\documentclass[10pt,conference,twocolumn]{IEEEtran}

\usepackage{amsmath}
\usepackage{amssymb}
\usepackage{latexsym}
\usepackage{amsfonts}

\begin{document}
\pagestyle{headings}  
\vspace{1cm}

%
\title{The $k$-error linear complexity distribution for $2^n$-periodic binary
sequences}

\author{
\authorblockN{Jianqin Zhou}
\authorblockA{ Telecommunication School, Hangzhou Dianzi University,
Hangzhou, 310018 China\\  Computer Science School, Anhui Univ. of
Technology, Ma'anshan, 243002 China\\ \ \ zhou9@yahoo.com\\
\ \\
Wanquan Liu\\
Department of Computing, Curtin University, Perth, WA 6102 Australia\\
 w.liu@curtin.edu.au
 }
}
\maketitle              

\begin{abstract}
The linear complexity and the $k$-error linear complexity of a
sequence have been used as important security measures for key stream sequence
strength in linear feedback shift register design. By studying the linear complexity of binary sequences with
period $2^n$, one could convert the computation of $k$-error
linear complexity into finding error sequences
with minimal Hamming weight. Based on Games-Chan algorithm,
the $k$-error linear complexity distribution of $2^n$-periodic binary
sequences is investigated in this paper. First, for $k=2,3$, the complete counting functions on the
$k$-error linear complexity of $2^n$-periodic balanced binary
sequences (with linear complexity less than $2^n$) are
characterized. Second, for $k=3,4$, the complete
counting functions on the $k$-error linear complexity of
$2^n$-periodic binary sequences with linear complexity $2^n$ are
presented. Third, as a consequence of these results, the counting functions
for the number of $2^n$-periodic binary sequences with the
$k$-error linear complexity for $k = 2$ and $3$ are obtained. Further
more, an important result in a recent paper is proved to be not
completely correct.  


\end{abstract}

\section{Introduction}

The linear complexity of a sequence is defined as the length of the shortest linear feedback shift register
(LFSR) that can generate the sequence.
The concept of linear complexity is very useful in the study of
security of stream ciphers for cryptographic applications and it has attracted many attentions in cryptographic community \cite{Ding,Stamp}.
In fact, a necessary condition for the security of a key stream generator in LFSR is
that it produces a sequence with high linear complexity. However,
high linear complexity can not necessarily guarantee the sequence is
safe since the linear complexity of some sequences is unstable. For example, if a
small number of changes to a sequence greatly reduce its linear
complexity, then the resulting key stream is cryptographically weak.
Ding, Xiao and Shan noticed this problem
first in their book \cite{Ding}, and proposed the weight complexity and sphere complexity.
Stamp and Martin \cite{Stamp} introduced $k$-error linear
complexity, which is similar to the sphere complexity, and put forward
the concept of $k$-error linear complexity profile. Specifically, suppose that $s$
is a sequence  with period $N$. For any $k(0\le k\le N)$,
$k$-error linear complexity of $s$, denoted as $L_k(s)$,  is defined
as the smallest linear complexity that can be obtained when any $k$
or fewer bits of the sequence are changed within one period.

One important result, proved by Kurosawa et al. in \cite{Kurosawa} is that the minimum number $k$
for which the $k$-error linear complexity of a $2^n$-periodic binary
sequence $s$ is strictly less than a linear complexity $L(s)$ of
$s$ is determined by $k_{\min}=2^{W_H(2^n-L(s))}$, where $W_H(a)$ denotes the Hamming weight of the
binary representation of an integer $a$. Also Rueppel \cite{Rueppel} derived  the number $N(L)$ of
$2^n$-periodic binary sequences with given linear complexity $L,
0\le L \le 2^n$.

For $k=1,2$, Meidl \cite{Meidl} characterized the complete
counting functions on the $k$-error linear complexity of
$2^n$-periodic binary sequences having maximal possible linear
complexity $2^n$. For $k=2,3$, Zhu and Qi \cite{Zhu} further showed
the complete counting functions on the $k$-error linear complexity
of $2^n$-periodic binary sequences with linear complexity $2^n-1$. By using algebraic and combinatorial methods, Fu et al. \cite{Fu} studied the linear complexity
and the $1$-error linear complexity for $2^n$-periodic binary sequences,
and characterized the
 $2^n$-periodic binary sequences with given 1-error linear
 complexity and derived the counting function for the 1-error
 linear complexity for $2^n$-periodic binary sequences.

By investigating sequences with linear complexity $2^n$ or linear complexity less than $2^n$ together, Kavuluru \cite{Kavuluru2008,Kavuluru} characterized $2^n$-periodic binary
sequences with fixed 2-error or 3-error linear complexity,  and obtained
the counting functions for the number of $2^n$-periodic binary
sequences with given $k$-error linear complexity for $k = 2$ and 3. These results are important progress on the $k$-error linear complexity.
Unfortunately, some of the results are not correct. In Section V, we prove that the counting functions in \cite{Kavuluru2008,Kavuluru} for the
number of $2^n$-periodic binary sequences with  3-error linear
complexity are incorrect in some cases.

In this paper, we will study the $k$-error linear complexity by proposing an approach different from those in current literature
\cite{Meidl,Zhu,Kavuluru2008,Kavuluru}. First we investigate sequences with  linear complexity $2^n$,  and sequences with linear complexity less than $2^n$, separately.  It is observed that for sequences with  linear complexity $2^n$, the $k$-error linear complexity is equal to  $(k+1)$-error linear complexity, when $k$ is odd. For sequences with  linear complexity less than $2^n$, the $k$-error linear complexity is equal to  $(k+1)$-error linear complexity, when $k$ is even. Then we investigate the $3$-error linear complexity in two cases and this reduces the complexity of this problem. Finally, by combining the results of two cases, we obtain the complete counting functions for the number
of $2^n$-periodic binary sequences with  $3$-error linear
complexity. The contribution of this paper can be summarized as follows. i) The results in \cite{Kavuluru2008,Kavuluru} on the $3$-error linear complexity are identified not completely correct and the correct results are given. It should be reminded that the $2$-error linear complexity was only solved partially in \cite{Zhu} and it was completely solved in \cite{Kavuluru}. Here we give a new complete solution in this paper as a by product. ii) A new approach is proposed for the $k$-error linear complexity problem, which can decompose this problem into two sub problems with less complexity.  iii) Generally, the complete counting functions for the number of $2^n$-periodic binary sequences with given $k$-error linear complexity for $k>3$ can be obtained using a similar approach.

The rest of this paper is organized as follows. In  Section II, some preliminary results are presented.
 In Section III, For $k=2,3$, the  counting functions on the $k$-error linear
complexity of $2^n$-periodic balanced binary sequences (with linear
complexity less than $2^n$) are  characterized.
In Section IV, for $k=3,4$, the  counting functions on the $k$-error linear complexity of $2^n$-periodic binary sequences with linear complexity  $2^n$ are derived.
 In Section V, the complete counting functions for the number
of $2^n$-periodic binary sequences with  $k$-error linear
complexity for $k = 2$ and $3$ are obtained, and an important result
in \cite{Kavuluru} is proved to be not completely correct.
Finally conclusions are given in Section VI

\section{Preliminaries}

In this section we give some preliminary results which will be used in the sequel.

We will consider sequences over $GF(q)$, which is the finite field
of order $q$. Let $x=(x_1,x_2,\cdots,x_n)$ and
$y=(y_1,y_2,\cdots,y_n)$ be vectors over $GF(q)$. Then define
$x+y=(x_1+y_1,x_2+y_2,\cdots,x_n+y_n)$.

When $n=2m$, we define
$Left(x)=(x_1,x_2,\cdots,x_m)$ and $Right(x)=(x_{m+1},x_{m+2},\cdots,x_{2m})$.

The Hamming weight of an $N$-periodic sequence $s$ is defined as the
number of   nonzero elements in per period of $s$, denoted by
$W(s)$. Let $s^N$ be one period of $s$. If $N=2^n$, $s^N$ is also
denoted as $s^{(n)}$. Obviously, $W(s^{(n)})=W(s^N)=W(s)$. $supp(s)$
is defined as a set of   nonzero elements in per period of $s$.

The generating function of a sequence $s=\{s_0, s_1, s_2, s_3,
\cdots, \}$  is defined by $$s(x)=s_0+ s_1x+ s_2x^2+ s_3x^3+
\cdots=\sum\limits^\infty_{i=0}s_ix^i$$

The generating function of a finite sequence $s^N=\{s_0, s_1, s_2,
 \cdots, s_{N-1},\}$ is defined by $s^N(x)=s_0+ s_1x+ s_2x^2+
\cdots+s_{N-1}x^{N-1}$. If $s$ is a periodic sequence with the first
period $s^N$, then,
\begin{eqnarray}
s(x) &=& s^N(x)(1+ x^N+ x^{2N}+ \cdots)=\frac{s^N(x)}{1-x^N}\notag\\
&=&\frac{s^N(x)/\gcd(s^N(x),1-x^N)}{(1-x^N)/\gcd(s^N(x),1-x^N)}\notag\\
&=&\frac{g(x)}{f_s(x)}\label{formula01}
\end{eqnarray}
where $f_s(x)=(1-x^N)/\gcd(s^N(x),1-x^N),
g(x)=s^N(x)/\gcd(s^N(x),1-x^N)$.

Obviously, $\gcd(g(x),f_s(x))=1, \deg(g(x)<\deg(f_s(x)))$. $f_s(x)$
is called  the minimal polynomial of $s$, and the degree of $f_s(x)$
is called the linear complexity of $s$, that is $\deg(f_s(x))=L(s)$.

Suppose that N=$2^n$ and $GF(q)=GF(2)$, then
$1-x^N=1-x^{2^n}=(1-x)^{2^n}=(1-x)^N$. Thus for binary sequences
with period $2^n$, its linear complexity is equal to the degree of
factor $(1-x)$ in $s^N(x)$.

The following three lemmas are  well known results on $2^n$-periodic
binary sequences.

\noindent {\bf Lemma  2.1} Suppose that s is a binary sequence with
period N=$2^n$, then L(s)=N if and only if the Hamming weight of a
period of the sequence is odd.

If an element one is removed from a sequence whose Hamming weight is
odd, the Hamming weight of the sequence will be changed to even, so
the main concern hereinafter is about sequences whose Hamming weight
are even.

\noindent {\bf Lemma 2.2}  Let $s_1$ and $s_2$ be two binary sequences
with period N=$2^n$. If $L(s_1)\ne L(s_2)$, then
$L(s_1+s_2)=\max\{L(s_1),L(s_2)\} $; otherwise if $L(s_1)= L(s_2)$,
then $L(s_1+s_2)<L(s_1)$.

Suppose that the linear complexity of s can decline when at least
$k$ elements of s are changed. By Lemma 2.2, the linear complexity
of the binary sequence, in which elements at exactly those $k$
positions are all nonzero, must be L(s). Therefore, for the
computation of $k$-error linear complexity, we only need to find the
binary sequence whose Hamming weight is minimum and its linear
complexity is L(s).

\noindent {\bf Lemma  2.3} Let $E_i$ be a $2^n$-periodic sequence
with one nonzero element at position $i$ and 0 elsewhere in each
period, $0\le i<2^n$. If $j-i=2^r(1+2a), a\ge0, 0\le i<j<2^n,
r\ge0$, then $L(E_i +E_j)=2^n-2^r$.

We now consider a binary sequence which has only 4 positions with
nonzero element.

\noindent {\bf Lemma  2.4} Let $s$ be a $2^n$-periodic sequence with
supp(s) $= \{i, j, k, l\}$, where $i<j,i<k<l$. If $d, e$ are the
largest integers for which $i\equiv j (\mod 2^d)$, $k\equiv l (\mod
2^e)$, and $k-i\equiv1 (\mod 2)$, then

$$L(s)=\left\{\begin{array}{l}
2^n-(1+2^d), \ \ \  \ \mbox{if}\ e=d\ \   \\
2^n-2^{\min(d,e)}, \ \ \ \ \mbox{otherwise}
\end{array}\right.$$

\begin{proof}\
According to Lemma 2.2, if $d\ne e$, then L(s)= $2^n-2^{\min(d,e)}$.

Consider the case of d=e. We know that $s$ can be decomposed into
the sum of two $E_{ij}$. The corresponding polynomial of $E_i+E_j$
is given by
\begin{eqnarray*}
x^i+x^j&=&x^i(1-x^{j-i})=x^i(1-x^{2^d(1+2u)})\\
&=&x^i(1-x^{2^d})(1+x^{2^d}+x^{2\cdot2^d}+\cdots +x^{2u\cdot2^d})
 \end{eqnarray*}

The corresponding polynomial of $E_k+E_l$ is given by
\begin{eqnarray*}
x^k+x^l&=&x^k(1-x^{l-k})=x^k(1-x^{2^d(1+2v)})\\
&=&x^k(1-x^{2^d})(1+x^{2^d}+x^{2\cdot2^d}+\cdots +x^{2v\cdot2^d})
 \end{eqnarray*}

Then $E_i +E_j+E_k +E_l$ corresponds to a polynomial, which is given
by
\begin{eqnarray*}
&&x^i+x^j+x^k+x^l\\
&=&x^i(1-x^{2^d})[(1+x^{2^d}+x^{2\cdot2^d}+\cdots +x^{2u\cdot2^d})\\
&&\ \ \ \ \ +x^{k-i}(1+x^{2^d}+x^{2\cdot2^d}+\cdots +x^{2v\cdot2^d})]\\
&=&x^i(1-x^{2^d})[1+x^{k-i}+(x^{2^d}+x^{2\cdot2^d}+\cdots +x^{2u\cdot2^d})\\
&&\ \ \ \ \ +x^{k-i}(x^{2^d}+x^{2\cdot2^d}+\cdots +x^{2v\cdot2^d})]\\
&=&x^i(1-x^{2^d})[1+x^{2c+1}+(x^{2^d}+x^{2\cdot2^d}+\cdots +x^{2u\cdot2^d})\\
&&\ \ \ \ \ +x^{k-i}(x^{2^d}+x^{2\cdot2^d}+\cdots +x^{2v\cdot2^d})]\\
&=&x^i(1-x)^{2^d+1}[(1+x+x^{2}+\cdots+x^{2c})\\
&&\ \ \  +(x^{2^d}+x^{3\cdot2^d}+\cdots +x^{(2u-1)\cdot2^d})(1+x)^{2^d-1}\\
&&\ \ \  +x^{k-i}(x^{2^d}+x^{3\cdot2^d}+\cdots
+x^{(2v-1)\cdot2^d})(1+x)^{2^d-1}]
 \end{eqnarray*}

There is no factor $(1+x)$ in $(1+x+x^{2}+\cdots+x^{2c})$, hence
$\gcd((1-x)^{2^n},x^i+x^j+x^k+x^l)=(1-x)^{2^d+1}$, thus,
$L(s)=2^n-(2^d+1)$.

This completes the proof.
\end{proof}\

\

\section{Counting functions with
the $2$-error linear complexity}

For $2^n$-periodic binary sequences with  linear complexity less than $2^n$,
 the change of one
bits per period results in a sequence with odd number of nonzero
bits per period, which has again linear complexity $2^n$. In this
section, we thus focus on the $2$-error linear complexity.

Further more, in order to derive the  counting functions
on the $3$-error linear complexity of $2^n$-periodic binary
sequences with linear complexity less than $2^n$, we only need to investigate the $2$-error linear complexity of $2^n$-periodic binary
sequences with linear complexity less than $2^n$.

Given a $2^n$-periodic binary sequence s, its linear complexity L(s)
can be  determined by the Games-Chan algorithm \cite{Games}. Based
on Games-Chan algorithm, the following  Lemma  3.1 is given in
\cite{Meidl2005}.

\noindent {\bf Lemma  3.1} Suppose that s is a binary sequence with
first period $s^{(n)}=\{s_0,s_1,s_2,\cdots, s_{2^n-1}\}$, a mapping
$\varphi_n$ from $F^{2^n}_2$ to $F^{2^{n-1}}_2$ is given by
\begin{eqnarray*}&&\varphi_n(s^{(n)})\\
&=&\varphi_n((s_0,s_1,s_2,\cdots,
s_{2^n-1}))\\
&=&(s_0+s_{2^{n-1}},s_1+s_{2^{n-1}+1},\cdots,
s_{2^{n-1}-1}+s_{2^n-1}) \end{eqnarray*}

Let $W(\mathbf{\upsilon})$ denote the Hamming weight of a vector
$\mathbf{\upsilon}$. Then mapping $\varphi_n$ has the following
properties

1) $W(\varphi_n(s^{(n)}))\le W(s^{(n)})$;

2) If $n\ge2$ then  $W(\varphi_n(s^{(n)}))$ and $W(s^{(n)})$ are
either both odd or both even;

3) The set $$\varphi^{-1}_{n+1}(s^{(n)})=\{v\in
F^{2^{n+1}}_2|\varphi_{n+1}(v)=s^{(n)} \}$$ of the preimage of
$s^{(n)}$ has cardinality $2^{2^n}$.

Rueppel \cite{Rueppel} presented the following.

\noindent {\bf Lemma  3.2}  The number $N(L)$ of $2^n$-periodic
binary sequences with given linear complexity $L, 0\le L \le 2^n$,
is given by $N(L)=\left\{\begin{array}{l}
1, \ \ \ \ \ L=0\ \   \\
2^{L-1}, \ 1\le L\le 2^n
\end{array}\right.$\

\

It is known that the computation of $k$-error linear complexity can be converted
to finding error sequences with minimal Hamming weight. Hence  2-error linear complexity of $s^{(n)}$ is  the smallest
linear complexity that can be obtained when any $u^{(n)}$ with
$W(u^{(n)})=0$ or 2 is added to $s^{(n)}$. So, the main approach of this section and next section is as follows. Let $s^{(n)}$ be a   binary sequence
with  linear complexity $c$, $u^{(n)}$  a  binary sequence with $W(u^{(n)})\le k$. We derive the  counting functions
on the $k$-error linear complexity of $2^n$-periodic binary
sequences by investigating $s^{(n)}+u^{(n)}$.
Based on this idea, we first prove the following lemmas

\noindent {\bf Lemma  3.3}  1). If $s^{(n)}$ is a   binary sequence
with  linear complexity $c, 1\le c\le 2^{n-1}-3$, $c\ne 2^{n-1}-2^m,
2\le m<n-1$,  $u^{(n)}$ is a  binary sequence, and $W(u^{(n)})=0$ or 2. Then the 2-error linear complexity
of $s^{(n)}+u^{(n)}$ is still $c$.

2). If $s^{(n)}$ is a   binary sequence with  linear complexity
$c=2^{n-1}-2^m, 0\le m<n-1$, then there exists a binary sequence
$u^{(n)}$ with $W(u^{(n)})=2$, such that the 2-error linear
complexity of $s^{(n)}+u^{(n)}$ is less than $c$.

\begin{proof}\
Without loose of  generality, we suppose that  $v^{(n)}\ne u^{(n)}$, and  $W(v^{(n)})=0$ or 2.

1). As $c\le 2^{n-1}-3$, we only need to consider the case
$L(u^{(n)}+v^{(n)})<2^{n-1}$. Thus
$Left(u^{(n)}+v^{(n)})=Right(u^{(n)}+v^{(n)})$ and
$W(Left(u^{(n)}+v^{(n)}))=2$.

 By Lemma 2.3, $L(u^{(n)}+v^{(n)})=2^{n-1}-2^m$, $0\le m<n-1$.

Thus $L(s^{(n)}+u^{(n)}+v^{(n)})\ge L(s^{(n)})$, so the 2-error
linear complexity of $s^{(n)}+u^{(n)}$ is  $c$.

2). As $s^{(n)}$ is a   binary sequence with  linear complexity
$c=2^{n-1}-2^m, 0\le m<n-1$. So the 2-error linear complexity of
$s^{(n)}+u^{(n)}$ must be less than $c$ when $L(u^{(n)}+v^{(n)})=c$.
\end{proof}\

\noindent {\bf Lemma  3.4} Suppose that $s^{(n)}$ and $t^{(n)}$ are
two different  binary sequences with  linear complexity $c, 1\le
c\le 2^{n-2}$, and   $u^{(n)}$ and $v^{(n)}$  are two different  binary
sequences, and
$W(u^{(n)})=0$ or 2, and $W(v^{(n)})=0$ or 2. Then $s^{(n)}+u^{(n)}\ne
t^{(n)}+v^{(n)}$.

\begin{proof}\ The following is obvious

$s^{(n)}+u^{(n)}\ne t^{(n)}+v^{(n)}$

$\Leftrightarrow$ $s^{(n)}+u^{(n)}+v^{(n)}\ne t^{(n)}$

$\Leftrightarrow$ $u^{(n)}+v^{(n)}\ne s^{(n)}+t^{(n)}$

Note that $s^{(n)}$ and $t^{(n)}$ are two different  binary
sequences with  linear complexity $c, 1\le c\le 2^{n-2}$, so the
linear complexity of $s^{(n)}+t^{(n)}$ is less than $2^{n-2}$, hence
one period of $s^{(n)}+t^{(n)}$ can be divided into 4 equal parts.

Suppose that $u^{(n)}+v^{(n)}= s^{(n)}+t^{(n)}$, then one period of $u^{(n)}+v^{(n)}$ can be divided into 4 equal parts.
 It follows that
the linear complexity of $u^{(n)}+v^{(n)}$ is  $2^{n-2}$, which
contradicts the fact that the linear complexity of $s^{(n)}+t^{(n)}$
is less than $2^{n-2}$.
\end{proof}\

Next we divide the 2-error linear complexity into three categories and deal with them respectively. First consider the category of $2^{n-1}-2^{n-m}$.

\noindent {\bf Lemma  3.5}  Let  $N_2(2^{n-1}-2^{n-m})$ be the
number of $2^n$-periodic binary sequences with  linear complexity less than
$2^n$ and given 2-error linear complexity $2^{n-1}-2^{n-m}, n\ge2,
1<m\le n$. Then
$$N_2(2^{n-1}-2^{n-m})=(1+\left(\begin{array}{c}2^n\\2\end{array}\right)-3\times2^{n+m-3}) 2^{2^{n-1}-2^{n-m}-1}
$$
\begin{proof}\
Suppose that $s^{(n)}$ is a   binary sequence with  linear
complexity $2^{n-1}-2^{n-m}$, and $u^{(n)}$ is a binary sequence
with  $W(u^{(n)})=2$. By Lemma  3.3, there exists a binary sequence $v^{(n)}$  with
 $W(v^{(n)})=2$, such that
$L(u^{(n)}+v^{(n)})=2^{n-1}-2^{n-m}$. So the 2-error linear
complexity of $u^{(n)}+s^{(n)}$ is less than $2^{n-1}-2^{n-m}$.

Suppose that $u^{(n)}$ is a binary sequence with linear complexity
$2^{n}$ and $W(u^{(n)})=2$, and there exist 2 nonzero elements whose
distance is  $2^{n-m}(2k+1)$ or $2^{n-1}$, with $k$ being an integer. It is
easy to verify that there exists a binary sequence $v^{(n)}$ with
 $W(v^{(n)})=2$, such that
$L(u^{(n)}+v^{(n)})=2^{n-1}-2^{n-m}$. So the 2-error linear
complexity of $u^{(n)}+s^{(n)}$ is less than $2^{n-1}-2^{n-m}$.

Let us divide one period of $u^{(n)}$   into $2^{n-m}$ subsequences
of form $\{a,a+2^{n-m}, a+2^{n-m+1},\cdots,
a+(2^m-1)\times2^{n-m}\}$.

If  2 nonzero elements of $u^{(n)}$ are in the same subsequence,
then the number of these $u^{(n)}$ can be given by
$$C1=2^{n-m}\times\left(\begin{array}{c}2^{m}\\2\end{array}\right)\times2^{m}.$$

 Suppose that  2 nonzero elements of
$u^{(n)}$ are in the same subsequence, and the distance of the 2
nonzero elements is not $2^{n-m}(2k+1)$, then the number of these
$u^{(n)}$ can be given by
$2^{n-m+1}\times\left(\begin{array}{c}2^{m-1}\\2\end{array}\right).$
Of  these $u^{(n)}$, there are
$2^{n-m}\times2^{m-1}=2^{n-1}$
sequences, in each sequence the distance of the 2 nonzero elements is
$2^{n-1}$.

So, if  2 nonzero elements of
$u^{(n)}$ are in the same subsequence, and the distance of the 2
nonzero elements is neither $2^{n-m}(2k+1)$ nor $2^{n-1}$, then the
number of these $u^{(n)}$ can be given by
$C2=2^{n-m+1}\times\left(\begin{array}{c}2^{m-1}\\2\end{array}\right)-2^{n-1}.$

Suppose that $u^{(n)}$ is a binary sequence with  $W(u^{(n)})=2$, and there exist  2 nonzero elements
whose distance is a multiple of $2^{n-m+1}$. Then there exists  one
binary sequence $v^{(n)}$  with
$W(v^{(n)})=2$, such that $L(u^{(n)}+v^{(n)})=2^{n-1}-2^{n-r},
1<r<m$. Let $t^{(n)}=s^{(n)}+u^{(n)}+v^{(n)}$. Then
$L(t^{(n)})=L(s^{(n)})=2^{n-1}-2^{n-m}$ and
$s^{(n)}+u^{(n)}=t^{(n)}+v^{(n)}$.

By Lemma  3.2,  the number  of $2^n$-periodic binary sequences with
given linear complexity $2^{n-1}-2^{n-m}$ is
$2^{2^{n-1}-2^{n-m}-1}$. This leads to the following,

{\scriptsize
\begin{eqnarray*}&&N_2(2^{n-1}-2^{n-m})\\
&=&[ 1+\left(\begin{array}{c}2^n\\2\end{array}\right)-(C1-C2)-C2/2]2^{2^{n-1}-2^{n-m}-1}\\
 &=&[
1+\left(\begin{array}{c}2^n\\2\end{array}\right)-2^{n-m}\left(\begin{array}{c}2^m\\2\end{array}\right)+2^{n-m}\left(\begin{array}{c}2^{m-1}\\2\end{array}\right)-2^{n-2}]\\
&&\times 2^{2^{n-1}-2^{n-m}-1}\\
&=&(1+\left(\begin{array}{c}2^n\\2\end{array}\right)-3\times2^{n+m-3}) 2^{2^{n-1}-2^{n-m}-1}\end{eqnarray*}}
\end{proof}\

Next we consider the category of $2^{n-1}-2^{n-m}+x$.

\noindent {\bf Lemma  3.6}  Let  $N_2(2^{n-1}-2^{n-m}+x)$ be the
number of $2^n$-periodic binary sequences with  linear complexity less than
$2^n$ and given 2-error linear complexity $2^{n-1}-2^{n-m}+x, n>3,
1<m<n-1, 0<x<2^{n-m-1}$.
Then{\small
\begin{eqnarray*}&&N_2(2^{n-1}-2^{n-m}+x)\\
&=&[1+\left(\begin{array}{c}2^n\\2\end{array}\right)+2^{r-m}-2^{r+m-2}]2^{2^{n-1}-2^{n-m}+x-1}\end{eqnarray*}}
\begin{proof}\
Suppose that $s^{(n)}$ is a binary sequence with linear complexity
$2^{n-1}-2^{n-m}+x$, and $u^{(n)}$ is a binary sequence with  $W(u^{(n)})=2$. By Lemma  3.3,  the
2-error linear complexity of $u^{(n)}+s^{(n)}$ is still
$2^{n-1}-2^{n-m}+x$.  The number of these $u^{(n)}$ can be given by
$\left(\begin{array}{c}2^{n}\\2\end{array}\right).$

Suppose that $u^{(n)}$ is a binary sequence with  $W(u^{(n)})=2$,  and there exist  2 nonzero elements
whose distance is  $2^{n-r}(1+2a), 1<r\le m, a\ge0$. Then there
exists one binary sequence $v^{(n)}$  with  $W(v^{(n)})=2$, such that $L(u^{(n)}+v^{(n)})=2^{n-1}-2^{n-r}$.
Let $t^{(n)}=s^{(n)}+u^{(n)}+v^{(n)}$. Then
$L(t^{(n)})=L(s^{(n)})=2^{n-1}-2^{n-m}+x$ and
$s^{(n)}+u^{(n)}=t^{(n)}+v^{(n)}$.

Let us divide one period of $u^{(n)}$   into $2^{n-m}$ subsequences
of form $\{a,a+2^{n-m}, a+2^{n-m+1},\cdots,
a+(2^m-1)\times2^{n-m}\}$.

If  2 nonzero elements of $u^{(n)}$ are in the same
subsequence, and their distance is
 $2^{n-1}$,  then there
exist $2^{m-1}-1$ binary sequences $v^{(n)}$ with $W(v^{(n)})=2$, such that
$L(u^{(n)}+v^{(n)})=2^{n-1}-2^{n-r}, 1<r\le m$. Let
$t^{(n)}=s^{(n)}+u^{(n)}+v^{(n)}$. Then
$s^{(n)}+u^{(n)}=t^{(n)}+v^{(n)}$. The number of these $u^{(n)}$ can
be given by
$D1=2^{n-m}\times2^{m-1}=2^{n-1}.$

Suppose that 2 nonzero elements of $u^{(n)}$ are in the same
subsequence, and their distance is not  $2^{n-1}$.
 Then there exist  one binary sequence $v^{(n)}$,  with $W(v^{(n)})=2$, such that
$L(u^{(n)}+v^{(n)})=2^{n-1}-2^{n-r}, 1<r\le m$.  The number of these $u^{(n)}$
can be given by
$$D2=2^{n-m}[\left(\begin{array}{c}2^{m}\\2\end{array}\right)-2^{m-1}]$$

By Lemma  3.2,  the number  of $2^n$-periodic binary sequences with
given linear complexity $2^{n-1}-2^{n-m}+x$ is
$2^{2^{n-1}-2^{n-m}+x-1}$. This will derive the following,
{\small
\begin{eqnarray*}&&N_2(2^{n-1}-2^{n-m}+x)\\
&=&[1+ \left(\begin{array}{c}2^n\\2\end{array}\right)- \frac{2^{m-1}-1}{2^{m-1}}\times D1-\frac{1}{2}\times D2]\\
&&\ \ \ \ 2^{2^{n-1}-2^{n-m}+x-1}\\
&=&\{1+ \left(\begin{array}{c}2^n\\2\end{array}\right)- \frac{2^{m-1}-1}{2^{m-1}}\times 2^{n-1}\\
&&-2^{n-m-1}[\left(\begin{array}{c}2^m\\2\end{array}\right)-2^{m-1}]\}2^{2^{n-1}-2^{n-m}+x-1}\\
&=&\{1+\left(\begin{array}{c}2^n\\2\end{array}\right)-(2^{m-1}-1)\times2^{n-m}\\
&&-2^{n-m-1}[\left(\begin{array}{c}2^{m}\\2\end{array}\right)-2^{m-1}]\}2^{2^{n-1}-2^{n-m}+x-1}\\
&=&[1+\left(\begin{array}{c}2^n\\2\end{array}\right)+2^{r-m}-2^{r+m-2}]2^{2^{n-1}-2^{n-m}+x-1}
\end{eqnarray*}}
\end{proof}\

\

Finally we consider the simplest category, that is $1\le
c\le 2^{r-2}-1$.

\noindent {\bf Lemma  3.7}  Let $L(r,c)=2^n-2^r+c, 3\le r\le n, 1\le
c\le 2^{r-2}-1$, and $N_3(L(r,c))$ be the number of $2^n$-periodic
binary sequences with  linear complexity less than $2^n$ and given 2-error
linear complexity $L(r,c)$. Then $$N_2(L)=\left\{\begin{array}{l}
1+\left(\begin{array}{c}2^n\\2\end{array}\right), \ \ \ \ \ \ \  \ L=0\ \   \\
2^{L-1}(1+\left(\begin{array}{c}2^r\\2\end{array}\right)), \
L=L(r,c)
\end{array}\right.$$
\begin{proof}\
Suppose that $s$ is a binary sequence with first period
$s^{(n)}=\{s_0,s_1,s_2,\cdots, s_{2^n-1}\}$, and $L(s)<2^n$. By
Games-Chan algorithm, $Left(s^{(t)})\ne Right(s^{(t)}), 1\le t\le n$,
 where
$s^{(t)}=\varphi_{t+1}\cdots\varphi_{n}(s^{(n)})$.

First consider the  case of $W(s^{(n)})=0$. There is only one
binary sequence of this kind.

Consider the  case of $W(s^{(n)})=2$. There is 2 nonzero bits in
$\{s_0,s_1,\cdots, s_{2^n-1}\}$, thus there are
$\left(\begin{array}{c}2^n\\2\end{array}\right)$ binary sequences of
this kind.

So $N_2(0)=1+\left(\begin{array}{c}2^n\\2\end{array}\right)$.

Consider $L(r,c)=2^n-2^r+c$, $3\le r\le n, 1\le c\le 2^{r-2}-1$.
Suppose that $s^{(n)}$ is a binary sequence with
$L(s^{(n)})=L(r,c)$. Note that
$L(r,c)=2^n-2^r+c=2^{n-1}+\cdots+2^r+c$.  By Games-Chan algorithm,
$Left(s^{(r)})= Right(s^{(r)})$, and $L(s^{(r)})=c$.

It is known that the number of  binary sequences $t^{(r)}$ with
$W(t^{(r)})=0$ or 2  is
$1+\left(\begin{array}{c}2^r\\2\end{array}\right)$.

By Lemma 3.3, the 2-error linear complexity of $s^{(r)}+t^{(r)}$ is
$c$.

By Lemma 3.2 and Lemma 3.4, the number of  binary sequences
$s^{(r)}+t^{(r)}$ is $2^{c-1}\times
(1+\left(\begin{array}{c}2^r\\2\end{array}\right))$

By Lemma 3.1, there are $2^{2^{n-1}+\cdots+2^r}=2^{2^{n}-2^r}$
binary sequences $s^{(n)}+t^{(n)}$,  such that
$s^{(r)}+t^{(r)}=\varphi_{r+1}\cdots\varphi_{n}(s^{(n)}+t^{(n)})$,
$t^{(r)}=\varphi_{r+1}\cdots\varphi_{n}(t^{(n)})$ and
$W(t^{(n)})=W(t^{(r)})$.

 Thus  the 2-error linear complexity of
$s^{(n)}+t^{(n)}$ is
$$2^{2^{n-1}+\cdots+2^r}+L_3(s^{(r)}+t^{(r)})=2^{2^{n}-2^r}+c=L(r,c).$$

Therefore, $N_2(L(r,c))=2^{2^{n}-2^r}\times 2^{c-1}\times
(1+\left(\begin{array}{c}2^r\\2\end{array}\right))=2^{L(r,c)-1}(1+\left(\begin{array}{c}2^r\\2\end{array}\right))$
\end{proof}\

Based on the results above, we have the following theorem.

 \noindent {\bf Theorem  3.1}  Let $L(r,c)=2^n-2^r+c$,  $2\le r\le
n, 1\le c\le 2^{r-1}-1$, and $N_2(L(r,c))$ be the number of
$2^n$-periodic binary sequences with  linear complexity less than
$2^n$ and given 2-error linear complexity $L(r,c)$.
 Then

{\scriptsize

$N_2(L)=\left\{\begin{array}{l}
\left(\begin{array}{c}2^{n}\\2\end{array}\right)+1, \ \ \ \ \ \ \ \ \ \ L=0\ \   \\
2^{L-1}(\left(\begin{array}{c}2^{r}\\2\end{array}\right)+1), \ L=L(r,c), 1\le c\le 2^{r-2}-1, r>2\\
2^{L-1}(\left(\begin{array}{c}2^{r}\\2\end{array}\right)+1-3\times2^{r+m-3}),  \\
 \ \ \ \ \  \ \ \ \ \ \ \ \ \ L=L(r,c),  c= 2^{r-1}-2^{r-m},  1<m\le r,r\ge2\\
2^{L-1}(\left(\begin{array}{c}2^{r}\\2\end{array}\right)+1+2^{r-m}-2^{r+m-2}),  \\
 \ \ \ \ \  \ \ \ \ \ \ \ \  L=L(r,c),  c= 2^{r-1}-2^{r-m}+x, \\
\ \ \ \ \  \ \ \ \ \ \ \ \   1<m<r-1, 0<x<2^{r-m-1}, r>3\\
0, \ \ \ \ \ \ \  \ \ \ \ \ \ \ \ \ \  \ \ \ \ \ \ \ \ \
\mbox{others}
\end{array}\right.$}
\begin{proof}\

By Lemma 3.7, we now only need to consider the case of $3\le r\le n, 2^{r-2}\le
c\le 2^{r-1}-1$.

By Lemma 3.1 and Lemma 3.5, $$N_2(L(r,c))=2^{L(r,c)-1}(\left(\begin{array}{c}2^{r}\\2\end{array}\right)+1-3\times2^{r+m-3})$$ for
$3\le r\le n,   c= 2^{r-1}-2^{r-m}, 1<m\le r$

By Lemma 3.1 and Lemma 3.6, $$N_2(L(r,c))=2^{L(r,c)-1}(\left(\begin{array}{c}2^{r}\\2\end{array}\right)+1+2^{r-m}-2^{r+m-2})$$ for
$4\le r\le n,   c= 2^{r-1}-2^{r-m}+x,1<m<r-1,0<x<2^{r-m-1}$

This completes the proof.
\end{proof}\

\

Now we give an example to illustrate Theorem 3.1.

For $n = 4$, the number of $2^n$-periodic binary sequences with
 linear complexity less than $2^n$ is $count = 2^{2^4-1} = 2^{15}$.

$N_2(L(2,1))=2^{12}$

$N_2(L(3,1))=2^8[\left(\begin{array}{c}2^{3}\\2\end{array}\right)+1]=count\times\frac{29}{128}$.

$N_2(L(3,2))=count\times\frac{17}{64}$.

$N_2(L(3,3))=count\times\frac{5}{32}$.

$N_2(0)=N_2(L(4,1))=\left(\begin{array}{c}2^{4}\\2\end{array}\right)+1=121$.

$N_2(L(4,2))=2\times121=242$.

$N_2(L(4,3))=4\times121=484$.

 $N_2(L(4,4))=776$.\ \ \
 $N_2(L(4,5))=1744$.

$N_2(L(4,6))=2336$.\ \ $N_2(L(4,7))=1600$.

It is easy to verify that the number of all these sequences is
$2^{15}$. These results are also checked by computer.

\

Notice that for $2^n$-periodic binary sequences with  linear complexity less than $2^n$,
 the change of three
bits per period results in a sequence with odd number of nonzero
bits per period, which has again linear complexity $2^n$. So from Theorem  3.1, we also know the counting functions on
the $3$-error linear complexity for $2^n$-periodic binary sequences with  linear complexity less than $2^n$.

\section{Counting functions on
the $3$-error linear complexity}

For $2^n$-periodic binary sequences with  linear complexity $2^n$,
 the change of two
bits per period results in a sequence with odd number of nonzero
bits per period, which has again linear complexity $2^n$. In this
section, we thus focus on the $3$-error linear complexity, which is
much more complicated than $1$-error linear complexity in \cite{Meidl2005}.

With an approach similar to that of previous section, we first investigate $s^{(n)}+u^{(n)}$.

\noindent {\bf Lemma  4.1} 1). If $s^{(n)}$ is a   binary sequence
with  linear complexity $c, 1\le c\le 2^{n-1}-3$, $c\ne 2^{n-1}-2^m,
2\le m<n-1$,  $u^{(n)}$ is a  binary sequence with linear complexity
$2^n$, and $W(u^{(n)})=1$ or 3. Then the 3-error linear complexity
of $s^{(n)}+u^{(n)}$ is also $c$.

2). If $s^{(n)}$ is a   binary sequence with  linear complexity
$c=2^{n-1}-2^m, 0\le m<n-1$, then there exists a binary sequence
$u^{(n)}$ with linear complexity $2^n$, such that the 3-error linear
complexity of $s^{(n)}+u^{(n)}$ is less than $c$.

\begin{proof}\
Note that 3-error linear complexity of $s^{(n)}$ is  the smallest
linear complexity that can be obtained when any $u^{(n)}$ with
$W(u^{(n)})=1$ or 3 is added to $s^{(n)}$.

Suppose that  $v^{(n)}\ne u^{(n)}$, and  $W(v^{(n)})=1$ or 3.

1). As $c\le 2^{n-1}-3$, we only need to consider the case
$L(u^{(n)}+v^{(n)})<2^{n-1}$. Thus
$Left(u^{(n)}+v^{(n)})=Right(u^{(n)}+v^{(n)})$ and
$W(Left(u^{(n)}+v^{(n)}))=2$.

 By Lemma 2.3, $L(u^{(n)}+v^{(n)})=2^{n-1}-2^m$, $0\le m<n-1$.

Thus $L(s^{(n)}+u^{(n)}+v^{(n)})\ge L(s^{(n)})$, so the 3-error
linear complexity of $s^{(n)}+u^{(n)}$ is  $c$.

2). As $s^{(n)}$ is a   binary sequence with  linear complexity
$c=2^{n-1}-2^m, 0\le m<n-1$. So the 3-error linear complexity of
$s^{(n)}+u^{(n)}$ must be less than $c$ when $L(u^{(n)}+v^{(n)})=c$.
\end{proof}\

\noindent {\bf Lemma  4.2} Suppose that $s^{(n)}$ and $t^{(n)}$ are
two different  binary sequences with  linear complexity $c, 1\le
c\le 2^{n-2}$, $u^{(n)}$ and $v^{(n)}$ are two different  binary
sequences with  linear complexity $2^n$, and $W(u^{(n)})=1$ or 3,
$W(v^{(n)})=1$ or 3. Then $s^{(n)}+u^{(n)}\ne t^{(n)}+v^{(n)}$.

\begin{proof}\ The following is obvious

$s^{(n)}+u^{(n)}\ne t^{(n)}+v^{(n)}$

$\Leftrightarrow$ $s^{(n)}+u^{(n)}+v^{(n)}\ne t^{(n)}$

$\Leftrightarrow$ $u^{(n)}+v^{(n)}\ne s^{(n)}+t^{(n)}$

Note that $s^{(n)}$ and $t^{(n)}$ are two different  binary
sequences with  linear complexity $c, 1\le c\le 2^{n-2}$, so the
linear complexity of $s^{(n)}+t^{(n)}$ is less than $2^{n-2}$, and
one period of $s^{(n)}+t^{(n)}$ can be divided into 4 equal parts.

Suppose that $u^{(n)}+v^{(n)}= s^{(n)}+t^{(n)}$, then one period of
$u^{(n)}+v^{(n)}$ can be divided into 4 equal parts, thus
$W(u^{(n)}+v^{(n)})=4$. It follows that the linear complexity of
$u^{(n)}+v^{(n)}$ is $2^{n-2}$, which contradicts with the fact that the
linear complexity of $s^{(n)}+t^{(n)}$ is less than $2^{n-2}$.
\end{proof}\

We divide 3-error linear complexity into three categories and deal with them respectively. First consider the category of $2^{n-1}-2^{n-m}$.

\noindent {\bf Lemma  4.3}  Let  $N_3(2^{n-1}-2^{n-m})$ be the
number of $2^n$-periodic binary sequences with  linear complexity
$2^n$ and given 3-error linear complexity $2^{n-1}-2^{n-m}, n>3,
1<m\le n$. Then{\scriptsize
\begin{eqnarray*}&&N_3(2^{n-1}-2^{n-m})\\
&=&[
\left(\begin{array}{c}2^n\\3\end{array}\right)-2^{n-m}\left(\begin{array}{c}2^m\\2\end{array}\right)-\left(\begin{array}{c}2^{n-m}\\2\end{array}\right)\left(\begin{array}{c}2^m\\2\end{array}\right)2^{m+1}\\
&&+\left(\begin{array}{c}2^{n-m}\\2\end{array}\right)\times2^{2m}(2^{m-2}-1)+2^{n-m-1}\times\left(\begin{array}{c}2^{m-1}\\3\end{array}\right)\\
&&-2^{n-2}\times(2^{m-2}-1)] 2^{2^{n-1}-2^{n-m}-1}\end{eqnarray*}}
\begin{proof}\
Suppose that $s^{(n)}$ is a   binary sequence with  linear
complexity $2^{n-1}-2^{n-m}$, and $u^{(n)}$ is a binary sequence
with linear complexity $2^{n}$ and $W(u^{(n)})=1$. It is obvious
that there exists a binary sequence $v^{(n)}$  with linear
complexity $2^{n}$ and $W(v^{(n)})=3$, such that
$L(u^{(n)}+v^{(n)})=2^{n-1}-2^{n-m}$. So the 3-error linear
complexity of $u^{(n)}+s^{(n)}$ is less than $2^{n-1}-2^{n-m}$.

Suppose that $u^{(n)}$ is a binary sequence with linear complexity
$2^{n}$ and $W(u^{(n)})=3$, and there exist 2 nonzero elements whose
distance is  $2^{n-m}(2k+1)$ or $2^{n-1}$, with $k$ being an integer. It is
easy to verify that there exists a binary sequence $v^{(n)}$ with
linear complexity $2^{n}$ and $W(v^{(n)})=3$, such that
$L(u^{(n)}+v^{(n)})=2^{n-1}-2^{n-m}$. So the 3-error linear
complexity of $u^{(n)}+s^{(n)}$ is less than $2^{n-1}-2^{n-m}$.

Let us divide one period of $u^{(n)}$   into $2^{n-m}$ subsequences
of form $\{a,a+2^{n-m}, a+2^{n-m+1},\cdots,
a+(2^m-1)\times2^{n-m}\}$.

If only 2 nonzero elements of $u^{(n)}$ are in the same subsequence,
then the number of these $u^{(n)}$ can be given by
$E1=2\left(\begin{array}{c}2^{n-m}\\2\end{array}\right)\times\left(\begin{array}{c}2^{m}\\2\end{array}\right)\times2^{m}.$

 Suppose that only 2 nonzero elements of
$u^{(n)}$ are in the same subsequence, and the distance of the 2
nonzero elements is not $2^{n-m}(2k+1)$, then the number of these
$u^{(n)}$ can be given by
$2\left(\begin{array}{c}2^{n-m}\\2\end{array}\right)\times2\times\left(\begin{array}{c}2^{m-1}\\2\end{array}\right)\times2^{m}.$
Of  these $u^{(n)}$, there are
$2\left(\begin{array}{c}2^{n-m}\\2\end{array}\right)\times2^{m-1}\times2^{m}$
sequences, in which the distance of the 2 nonzero elements is
$2^{n-1}$.

 So, if only 2 nonzero elements of
$u^{(n)}$ are in the same subsequence, and the distance of the 2
nonzero elements is neither $2^{n-m}(2k+1)$ nor $2^{n-1}$, then the
number of these $u^{(n)}$ can be given by

\begin{eqnarray*}E2&=&2\left(\begin{array}{c}2^{n-m}\\2\end{array}\right)\times2\times\left(\begin{array}{c}2^{m-1}\\2\end{array}\right)\times2^{m}\\
&&-2\left(\begin{array}{c}2^{n-m}\\2\end{array}\right)\times2^{m-1}\times2^{m}\\
&=&\left(\begin{array}{c}2^{n-m}\\2\end{array}\right)\times2^{2m}\times(2^{m-1}-2)\end{eqnarray*}

Suppose that $u^{(n)}$ is a binary sequence with linear complexity
$2^{n}$ and $W(u^{(n)})=3$, and there exist  2 nonzero elements
whose distance is a multiple of $2^{n-m+1}$. Then there exists  one
binary sequence $v^{(n)}$  with linear complexity $2^{n}$ and
$W(v^{(n)})=3$, such that $L(u^{(n)}+v^{(n)})=2^{n-1}-2^{n-r},
1<r<m$. Let $t^{(n)}=s^{(n)}+u^{(n)}+v^{(n)}$. Then
$L(t^{(n)})=L(s^{(n)})=2^{n-1}-2^{n-m}$ and
$s^{(n)}+u^{(n)}=t^{(n)}+v^{(n)}$, where there exist exactly 2
nonzero elements in $v^{(n)}$ whose distance  is a multiple of
 $2^{n-m+1}$.

If all 3 nonzero elements of $u^{(n)}$ are in the same subsequence,
then the number of these $u^{(n)}$ can be given by
$E3=2^{n-m}\times\left(\begin{array}{c}2^{m}\\3\end{array}\right).$

 Suppose that all 3 nonzero elements of
$u^{(n)}$ are in the same subsequence, and there do not exist  2
nonzero elements whose distance is $2^{n-m}(2k+1)$, then the number
of these $u^{(n)}$ can be given by
$2^{n-m+1}\times\left(\begin{array}{c}2^{m-1}\\3\end{array}\right).$

Of  these $u^{(n)}$, there are
$2^{n-m+1}\times2^{m-2}\times(2^{m-1}-2)=2^{n}\times(2^{m-2}-1)$
sequences, in which there  exist  2 nonzero elements whose distance
is  $2^{n-1}$.

 So, if  all 3 nonzero elements of
$u^{(n)}$ are in the same subsequence, and there do not exist  2
nonzero elements whose distance is $2^{n-m}(2k+1)$ or $2^{n-1}$,
then the number of these $u^{(n)}$ can be given by
$$E4=2^{n-m+1}\times\left(\begin{array}{c}2^{m-1}\\3\end{array}\right)-2^{n}\times(2^{m-2}-1)$$

Suppose that $u^{(n)}$ is a binary sequence with linear complexity
$2^{n}$ and  all 3 nonzero elements of $u^{(n)}$ are in the same
subsequence, and there do not exist  2 nonzero elements whose
distance is $2^{n-m}(2k+1)$ or $2^{n-1}$.
 Then there exist  3 distinct
binary sequences $v^{(n)}_i,1\le i\le3,$  with linear complexity
$2^{n}$ and $W(v^{(n)}_i)=3$, such that
$L(u^{(n)}+v^{(n)}_i)=2^{n-1}-2^{n-r}, 1<r<m$. Let
$t^{(n)}_i=s^{(n)}+u^{(n)}+v^{(n)}_i$. Then
$L(t^{(n)}_i)=L(s^{(n)})=2^{n-1}-2^{n-m}$ and
$s^{(n)}+u^{(n)}=t^{(n)}_i+v^{(n)}_i$.

By Lemma  3.2,  the number  of $2^n$-periodic binary sequences with
given linear complexity $2^{n-1}-2^{n-m}$ is
$2^{2^{n-1}-2^{n-m}-1}$. It follows that,

{\scriptsize
\begin{eqnarray*}&&N_3(2^{n-1}-2^{n-m})\\
&=&[ \left(\begin{array}{c}2^n\\3\end{array}\right)-E3-E1+E2/2+E4/4] 2^{2^{n-1}-2^{n-m}-1}\\
 &=&[
\left(\begin{array}{c}2^n\\3\end{array}\right)-2^{n-m}\left(\begin{array}{c}2^m\\3\end{array}\right)-\left(\begin{array}{c}2^{n-m}\\2\end{array}\right)\left(\begin{array}{c}2^m\\2\end{array}\right)2^{m+1}\\
&&+\left(\begin{array}{c}2^{n-m}\\2\end{array}\right)\times2^{2m}(2^{m-2}-1)\\
&&+2^{n-m-1}\times\left(\begin{array}{c}2^{m-1}\\3\end{array}\right)-2^{n-2}\times(2^{m-2}-1)]
2^{2^{n-1}-2^{n-m}-1}\end{eqnarray*}}
\end{proof}\

For $m=2$,

{\small $$N_3(2^{n-2})=\{
\left(\begin{array}{c}2^n\\3\end{array}\right)-[2^n+2^{n-2}\left(\begin{array}{c}4\\2\end{array}\right)(2^n-4)]
\}2^{2^{n-2}-1}$$}

The following are examples.

For $n=3$,
$\left(\begin{array}{c}2^n\\3\end{array}\right)-[2^n+2^{n-2}\left(\begin{array}{c}4\\2\end{array}\right)(2^n-4)]=0$,
which means that there is no $2^3$-periodic binary sequences with
linear complexity $2^{3}$ and 3-error linear complexity 2.

For $n=4$, {\small $$\{
\left(\begin{array}{c}2^n\\3\end{array}\right)-[2^n+2^{n-2}\left(\begin{array}{c}4\\2\end{array}\right)(2^n-4)]
\}2^{2^{n-2}-1}=2048$$} which means that there is 2048
 binary sequences of period $2^4$ with linear complexity $2^{4}$ and
3-error linear complexity $2^2$.

For $m=3$,
{\scriptsize
$$N_3(2^{n-2}+2^{n-3})=[
\left(\begin{array}{c}2^n\\3\end{array}\right)-7\times2^n-384\left(\begin{array}{c}2^{n-3}\\2\end{array}\right)]
2^{2^{n-2}+2^{n-3}-1}$$}

For $m=4$,
{\scriptsize
$$N_3(2^{n-1}-2^{n-4})=[
\left(\begin{array}{c}2^n\\3\end{array}\right)-34\times2^n-3072\left(\begin{array}{c}2^{n-4}\\2\end{array}\right)]
2^{2^{n-1}-2^{n-4}-1}$$}

\

Next we consider the category of $2^{n-1}-2^{n-m}+x$.

\noindent {\bf Lemma  4.4}  Let  $N_3(2^{n-1}-2^{n-m}+x)$ be the
number of $2^n$-periodic binary sequences with  linear complexity
$2^n$ and given 3-error linear complexity $2^{n-1}-2^{n-m}+x, n>3,
1<m<n-1, 0<x<2^{n-m-1}$.
Then{\tiny
\begin{eqnarray*}&&N_3(2^{n-1}-2^{n-m}+x)\\
&=&\{
\left(\begin{array}{c}2^n\\3\end{array}\right)-(2^{m-2}-1)\times2^{n+1}-(2^{m-1}-1)\times\left(\begin{array}{c}2^{n-m}\\2\end{array}\right)\times2^{m+1}\\
&&-3\times2^{n-m-2}[\left(\begin{array}{c}2^m\\3\end{array}\right)-4\left(\begin{array}{c}2^{m-1}\\2\end{array}\right)]\\
&&-\left(\begin{array}{c}2^{n-m}\\2\end{array}\right)\times[\left(\begin{array}{c}2^m\\2\end{array}\right)-2^{m-1}]\times2^m\}
2^{2^{n-1}-2^{n-m}+x-1}\end{eqnarray*}}
\begin{proof}\
Suppose that $s^{(n)}$ is a binary sequence with linear complexity
$2^{n-1}-2^{n-m}+x$, and $u^{(n)}$ is a binary sequence with linear
complexity $2^{n}$ and $W(u^{(n)})=1$ or 3. By Lemma  4.1,  the
3-error linear complexity of $u^{(n)}+s^{(n)}$ is still
$2^{n-1}-2^{n-m}+x$.  The number of these $u^{(n)}$ can be given by
$\left(\begin{array}{c}2^{n}\\3\end{array}\right) +2^n.$

Suppose that $u^{(n)}$ is a binary sequence with linear complexity
$2^{n}$ and $W(u^{(n)})=1$. It is easy to show that

 1
binary sequence $v^{(n)}$ with linear complexity $2^{n}$ and
$W(v^{(n)})=3$, such that $L(u^{(n)}+v^{(n)})=2^{n-1}-2^{n-2}$.

2 binary sequences $v^{(n)}$ with linear complexity $2^{n}$ and
$W(v^{(n)})=3$, such that $L(u^{(n)}+v^{(n)})=2^{n-1}-2^{n-3}$.

$\cdots\cdots$

$2^{m-2}$ binary sequences $v^{(n)}$ with linear complexity $2^{n}$
and $W(v^{(n)})=3$, such that $L(u^{(n)}+v^{(n)})=2^{n-1}-2^{n-m}$.

Let $t^{(n)}=s^{(n)}+u^{(n)}+v^{(n)}$. Then
$L(t^{(n)})=L(s^{(n)})=2^{n-1}-2^{n-m}+x$ and
$s^{(n)}+u^{(n)}=t^{(n)}+v^{(n)}$, note that there are $2^{m-1}-1$
such  $t^{(n)}$ and $v^{(n)}$. Thus we will not consider these
$u^{(n)}$ with $W(u^{(n)})=1$.

Suppose that $u^{(n)}$ is a binary sequence with linear complexity
$2^{n}$ and $W(u^{(n)})=3$, and there exist  2 nonzero elements
whose distance is  $2^{n-r}(1+2a), 1<r\le m, a\ge0$. Then there
exists one binary sequence $v^{(n)}$  with linear complexity $2^{n}$
and $W(v^{(n)})=3$, such that $L(u^{(n)}+v^{(n)})=2^{n-1}-2^{n-r}$.
Let $t^{(n)}=s^{(n)}+u^{(n)}+v^{(n)}$. Then
$L(t^{(n)})=L(s^{(n)})=2^{n-1}-2^{n-m}+x$ and
$s^{(n)}+u^{(n)}=t^{(n)}+v^{(n)}$.

Let us divide one period of $u^{(n)}$   into $2^{n-m}$ subsequences
of form $\{a,a+2^{n-m}, a+2^{n-m+1},\cdots,
a+(2^m-1)\times2^{n-m}\}$.

If  all 3 nonzero elements of $u^{(n)}$ are in the same
subsequence, and there  exist  2 nonzero elements whose distance is
 $2^{n-1}$,  then there
exist $2^{m-1}-2$ binary sequences $v^{(n)}$  with linear complexity
$2^{n}$ and $W(v^{(n)})=3$, such that
$L(u^{(n)}+v^{(n)})=2^{n-1}-2^{n-r}, 1<r\le m$. Let
$t^{(n)}=s^{(n)}+u^{(n)}+v^{(n)}$. Then
$s^{(n)}+u^{(n)}=t^{(n)}+v^{(n)}$. The number of these $u^{(n)}$ can
be given by
$F1=2^{n-m}\times2\times\left(\begin{array}{c}2^{m-1}\\2\end{array}\right)\times2.$

Suppose that $u^{(n)}$ is a binary sequence with linear complexity
$2^{n}$ and  all 3 nonzero elements of $u^{(n)}$ are in the same
subsequence, and there do not exist  2 nonzero elements whose
distance is  $2^{n-1}$.
 Then there exist  3 distinct
binary sequences $v^{(n)}_i,1\le i\le3,$  with linear complexity
$2^{n}$ and $W(v^{(n)}_i)=3$, such that
$L(u^{(n)}+v^{(n)}_i)=2^{n-1}-2^{n-r}, 1<r\le m$. Let
$t^{(n)}_i=s^{(n)}+u^{(n)}+v^{(n)}_i$. Then
$s^{(n)}+u^{(n)}=t^{(n)}_i+v^{(n)}_i$. The number of these $u^{(n)}$
can be given by
$$F2=2^{n-m}[\left(\begin{array}{c}2^{m}\\3\end{array}\right)-2\times\left(\begin{array}{c}2^{m-1}\\2\end{array}\right)\times2]$$

 If only 2 nonzero elements of
$u^{(n)}$ are in the same subsequence, and the distance of the 2
nonzero elements is  $2^{n-1}$, then there exist $2^{m-1}-1$ binary
sequences $v^{(n)}$  with linear complexity $2^{n}$ and
$W(v^{(n)})=3$, such that $L(u^{(n)}+v^{(n)})=2^{n-1}-2^{n-r},
1<r\le m$. Let $t^{(n)}=s^{(n)}+u^{(n)}+v^{(n)}$. Then
$s^{(n)}+u^{(n)}=t^{(n)}+v^{(n)}$. The number of these $u^{(n)}$ can
be given by
$$F3=2\left(\begin{array}{c}2^{n-m}\\2\end{array}\right)\times2^{m-1}\times2^{m}=\left(\begin{array}{c}2^{n-m}\\2\end{array}\right)\times2^{2m}.$$

If only 2 nonzero elements of $u^{(n)}$ are in the same
subsequence, and the distance of the 2 nonzero elements is not
$2^{n-1}$, then there exists one binary sequence $v^{(n)}$ with
linear complexity $2^{n}$ and $W(v^{(n)})=3$, such that
$L(u^{(n)}+v^{(n)})=2^{n-1}-2^{n-r}, 1<r\le m$. Let
$t^{(n)}=s^{(n)}+u^{(n)}+v^{(n)}$. Then
$s^{(n)}+u^{(n)}=t^{(n)}+v^{(n)}$. The number of these $u^{(n)}$ can
be given by
$$F4=2\left(\begin{array}{c}2^{n-m}\\2\end{array}\right)\times[\left(\begin{array}{c}2^{m}\\2\end{array}\right)-2^{m-1}]\times2^m$$

By Lemma  3.2,  the number  of $2^n$-periodic binary sequences with
given linear complexity $2^{n-1}-2^{n-m}+x$ is
$2^{2^{n-1}-2^{n-m}+x-1}$. It follows that,
{\tiny
\begin{eqnarray*}&&N_3(2^{n-1}-2^{n-m}+x)\\
&=&[ \left(\begin{array}{c}2^n\\3\end{array}\right)- \frac{2^{m-1}-2}{2^{m-1}-1}\times F1-\frac{2^{m-1}-1}{2^{m-1}}\times F3\\
&&-\frac{3}{4}\times F2-\frac{1}{2}\times F4]2^{2^{n-1}-2^{n-m}+x-1}\\
&=&\{ \left(\begin{array}{c}2^n\\3\end{array}\right)- \frac{2^{m-1}-2}{2^{m-1}-1}\times 2^{n-m+2}\left(\begin{array}{c}2^{m-1}\\2\end{array}\right)\\
&&-\frac{2^{m-1}-1}{2^{m-1}}\times \left(\begin{array}{c}2^{n-m}\\2\end{array}\right)\times2^{2m}\\
&&-\frac{3}{4}\times 2^{n-m}[\left(\begin{array}{c}2^m\\3\end{array}\right)-4\left(\begin{array}{c}2^{m-1}\\2\end{array}\right)]\\
&&-\frac{1}{2}\times 2\left(\begin{array}{c}2^{n-m}\\2\end{array}\right)\times[\left(\begin{array}{c}2^m\\2\end{array}\right)-2^{m-1}]\times2^m\}2^{2^{n-1}-2^{n-m}+x-1}\\
&=&\{
\left(\begin{array}{c}2^n\\3\end{array}\right)-(2^{m-2}-1)\times2^{n+1}-(2^{m-1}-1)\times\left(\begin{array}{c}2^{n-m}\\2\end{array}\right)\times2^{m+1}\\
&&-3\times2^{n-m-2}[\left(\begin{array}{c}2^m\\3\end{array}\right)-4\left(\begin{array}{c}2^{m-1}\\2\end{array}\right)]\\
&&-\left(\begin{array}{c}2^{n-m}\\2\end{array}\right)\times[\left(\begin{array}{c}2^m\\2\end{array}\right)-2^{m-1}]\times2^m\}
2^{2^{n-1}-2^{n-m}+x-1}\end{eqnarray*}}
\end{proof}\

A special case is given below.

For $m=2, x=1$,

$$N_3(2^{n-2}+1)=[
\left(\begin{array}{c}2^n\\3\end{array}\right)-2^{n-3}\left(\begin{array}{c}4\\2\end{array}\right)(2^n-4)]
2^{2^{n-2}}$$

\

Finally we consider the simplest category, that is $1\le
c\le 2^{r-2}-1$.

\noindent {\bf Lemma  4.5}  Let $L(r,c)=2^n-2^r+c, 3\le r\le n, 1\le
c\le 2^{r-2}-1$, and $N_3(L(r,c))$ be the number of $2^n$-periodic
binary sequences with  linear complexity $2^n$ and given 3-error
linear complexity $L(r,c)$. Then $$N_3(L)=\left\{\begin{array}{l}
\left(\begin{array}{c}2^n\\3\end{array}\right)+2^n, \ \ \ \ \ \ \  \ L=0\ \   \\
2^{L-1}(\left(\begin{array}{c}2^r\\3\end{array}\right)+2^r), \
L=L(r,c)
\end{array}\right.$$
\begin{proof}\
Suppose that $s$ is a binary sequence with first period
$s^{(n)}=\{s_0,s_1,s_2,\cdots, s_{2^n-1}\}$, and $L(s)=2^n$. By
Games-Chan algorithm, $Left(s^{(t)})\ne Right(s^{(t)}), 1\le t\le n,
L(s^{(0)})=1$, where
$s^{(t)}=\varphi_{t+1}\cdots\varphi_{n}(s^{(n)})$.

Thus $s^{(0)}=\{1\}$, and $s_0+s_1+\cdots+ s_{2^t-1}=1, 1\le t\le n$

First consider the  case of $W(s^{(n)})=1$. There is only one
nonzero bit in $\{s_0,s_1,\cdots, s_{2^n-1}\}$, thus there are $2^n$
binary sequences of this kind.

Consider the  case of $W(s^{(n)})=3$. There is 3 nonzero bits in
$\{s_0,s_1,\cdots, s_{2^n-1}\}$, thus there are
$\left(\begin{array}{c}2^n\\3\end{array}\right)$ binary sequences of
this kind.

So $N_3(0)=\left(\begin{array}{c}2^n\\3\end{array}\right)+2^n$.

Consider $L(r,c)=2^n-2^r+c$, $3\le r\le n, 1\le c\le 2^{r-2}-1$.
Suppose that $s^{(n)}$ is a binary sequence with
$L(s^{(n)})=L(r,c)$. Note that
$L(r,c)=2^n-2^r+c=2^{n-1}+\cdots+2^r+c$.  By Games-Chan algorithm,
$Left(s^{(r)})= Right(s^{(r)})$, and $L(s^{(r)})=c$.

It is known that the number of  binary sequences $t^{(r)}$ with
$W(t^{(r)})=1$ or 3 and $L(t^{(r)})=2^r$ is
$\left(\begin{array}{c}2^r\\3\end{array}\right)+2^r$.

By Lemma 4.1, the 3-error linear complexity of $s^{(r)}+t^{(r)}$ is
$c$.

By Lemma 3.2 and Lemma 4.2, the number of  binary sequences
$s^{(r)}+t^{(r)}$ is $2^{c-1}\times
(\left(\begin{array}{c}2^r\\3\end{array}\right)+2^r)$

By Lemma 3.1, there are $2^{2^{n-1}+\cdots+2^r}=2^{2^{n}-2^r}$
binary sequences $s^{(n)}+t^{(n)}$,  such that
$s^{(r)}+t^{(r)}=\varphi_{r+1}\cdots\varphi_{n}(s^{(n)}+t^{(n)})$,
$t^{(r)}=\varphi_{r+1}\cdots\varphi_{n}(t^{(n)})$ and
$W(t^{(n)})=W(t^{(r)})$.

 Thus  the 3-error linear complexity of
$s^{(n)}+t^{(n)}$ is
$$2^{2^{n-1}+\cdots+2^r}+L_3(s^{(r)}+t^{(r)})=2^{2^{n}-2^r}+c=L(r,c).$$

Therefore, $N_3(L(r,c))=2^{2^{n}-2^r}\times 2^{c-1}\times
(\left(\begin{array}{c}2^r\\3\end{array}\right)+2^r)=2^{L(r,c)-1}(\left(\begin{array}{c}2^r\\3\end{array}\right)+2^r)$
\end{proof}\

Based on the results above, we can have the following theorem.

 \noindent {\bf Theorem  4.1}  Let $L(r,c)=2^n-2^r+c$, or $2^n-2^3+1$, $4\le r\le
n, 1\le c\le 2^{r-1}-1$, and $N_3(L(r,c))$ be the number of
$2^n$-periodic binary sequences with  linear complexity $2^n$ and
given 3-error linear complexity $L(r,c)$. Let {\scriptsize
\begin{eqnarray*}&&f(n,m)\\
&=&
\left(\begin{array}{c}2^n\\3\end{array}\right)-2^{n-m}\left(\begin{array}{c}2^m\\3\end{array}\right)-\left(\begin{array}{c}2^{n-m}\\2\end{array}\right)\left(\begin{array}{c}2^m\\2\end{array}\right)2^{m+1}\\
&&+\left(\begin{array}{c}2^{n-m}\\2\end{array}\right)\times2^{2m}(2^{m-2}-1)+2^{n-m-1}\times\left(\begin{array}{c}2^{m-1}\\3\end{array}\right)\\
&&-2^{n-2}\times(2^{m-2}-1)\\
\ \\
&&g(n,m)\\
&=&
\left(\begin{array}{c}2^n\\3\end{array}\right)-(2^{m-2}-1)\times2^{n+1}\\
&&-(2^{m-1}-1)\times\left(\begin{array}{c}2^{n-m}\\2\end{array}\right)\times2^{m+1}\\
&&-3\times2^{n-m-2}[\left(\begin{array}{c}2^m\\3\end{array}\right)-4\left(\begin{array}{c}2^{m-1}\\2\end{array}\right)]\\
&&-\left(\begin{array}{c}2^{n-m}\\2\end{array}\right)\times[\left(\begin{array}{c}2^m\\2\end{array}\right)-2^{m-1}]\times2^m\end{eqnarray*}}

 Then

{\scriptsize $N_3(L)=\left\{\begin{array}{l}
\left(\begin{array}{c}2^{n}\\3\end{array}\right)+2^n,  \ \ \  \ L=0\ \   \\
2^{L(r,c)-1}(\left(\begin{array}{c}2^{r}\\3\end{array}\right)+2^r),\\
 \ \ \ \ \  \ \ \  L=L(r,c), 1\le c\le 2^{r-2}-1, r>2\\
2^{L(r,c)-1}f(r,m), \\
\ \ \ \ \ \ \ \ L=L(r,c),  c= 2^{r-1}-2^{r-m},1<m\le r, r>3\\
2^{L(r,c)-1}g(r,m),\\
 \ \ \ \ \  \ \ \  L=L(r,c),  c= 2^{r-1}-2^{r-m}+x,\\
 \ \ \ \ \  \ \ \    1<m<r-1,0<x<2^{r-m-1},r>3\\
0, \ \ \ \ \ \ \  \ \ \ \ \ \ \ \ \ \  \ \ \ \  \ \mbox{others}
\end{array}\right.$}

\begin{proof}\
For $n=3$, the number  of $2^n$-periodic binary sequences with given
linear complexity $2^{n}$ is  $2^{2^{3}-1}=128$.

By Lemma 4.5,
$N_3(0)=N_3(L(3,1))=\left(\begin{array}{c}2^{n}\\3\end{array}\right)+2^n=64$.
Thus,  $N_3(L)=0$ for $L\ne 0$ and $L\ne L(3,1)$.

By Lemma 4.5, we now only need to consider $4\le r\le n, 2^{r-2}\le
c\le 2^{r-1}-1$.

By Lemma 3.1 and Lemma 4.3, $N_3(L(r,c))=2^{L(r,c)-1}f(r,m)$ for
$4\le r\le n,   c= 2^{r-1}-2^{r-m}, 1<m\le r$

By Lemma 3.1 and Lemma 4.4, $N_3(L(r,c))=2^{L(r,c)-1}g(r,m)$ for
$4\le r\le n,   c= 2^{r-1}-2^{r-m}+x,1<m<r-1,0<x<2^{r-m-1}$
\end{proof}\

\

The following is an example to illustrate Theorem 4.1.

For $n=4$, the number  of $2^n$-periodic binary sequences with given
linear complexity $2^{n}$ is  $count=2^{2^{4}-1}=2^{15}$.

By Lemma 4.5,
$N_3(L(3,1))=2^8[\left(\begin{array}{c}2^{3}\\3\end{array}\right)+2^3]=count/2$.

$N_3(0)=N_3(L(4,1))=\left(\begin{array}{c}2^{4}\\3\end{array}\right)+2^4=count\times\frac{9}{512}$.

$N_3(L(4,2))=2\times
count\times\frac{9}{512}=count\times\frac{9}{256}$.

$N_3(L(4,3))=4\times
count\times\frac{9}{512}=count\times\frac{9}{128}$.

By Lemma 4.3, $N_3(L(4,4))=\{
\left(\begin{array}{c}2^n\\3\end{array}\right)-[2^n+2^{n-2}\left(\begin{array}{c}4\\2\end{array}\right)(2^n-4)]
\}2^{2^{n-2}-1}=\frac{count}{16}$.

By Lemma 4.4, $N_3(L(4,5))=[
\left(\begin{array}{c}2^n\\3\end{array}\right)-2^{n-3}\left(\begin{array}{c}4\\2\end{array}\right)(2^n-4)]
2^{2^{n-2}}=count\times\frac{13}{64}$.

By Lemma 4.3, $N_3(L(4,6))=[
\left(\begin{array}{c}2^n\\3\end{array}\right)-7\times2^n-384\left(\begin{array}{c}2^{n-3}\\2\end{array}\right)]
2^{2^{n-2}+2^{n-3}-1}=\frac{count}{16}$.

By Lemma 4.3, $N_3(L(4,7))=[
\left(\begin{array}{c}2^n\\3\end{array}\right)-34\times2^n-3072\left(\begin{array}{c}2^{n-4}\\2\end{array}\right)]
2^{2^{n-1}-2^{n-4}-1}=\frac{count}{32}$.

It is easy to verify that the number of all these sequences is
$2^{15}$. These results are also checked by computer.

\

\section{Complete counting functions on
the $2$-error or $3$-error linear complexity}

The following theorem is presented in \cite{Meidl2005}.

\noindent {\bf Theorem  5.1}  Let $L(r,c)=2^n-2^r+c,2\le r\le n,
1\le c\le 2^{r-1}-1$, and $N_1(L(r,c))$ be the number of
$2^n$-periodic binary sequences with  linear complexity $2^n$ and
given 1-error linear complexity $L(r,c)$. Then
$N_1(L)=\left\{\begin{array}{l}
2^n, \ \ \ \ \ \  \ L=0\ \   \\
2^{L+r-1}, \ L=L(r,c)\\
0, \ \ \ \ \ \ \  \ \ \mbox{others}
\end{array}\right.$

\

From Theorem  5.1, we have  counting functions on
the $2$-error  linear complexity for $2^n$-periodic binary sequences with
linear complexity $2^n$. From Theorem  3.1, we have  counting functions on
the $2$-error  linear complexity for $2^n$-periodic binary sequences with
linear complexity less than $2^n$.

 By combining the results of two cases, it  is easy to derive the complete counting functions for the number
of $2^n$-periodic binary sequences with fixed $2$-error linear
complexity.

 \noindent {\bf Theorem  5.2}  Let $L(r,c)=2^n-2^r+c$,  $2\le r\le
n, 1\le c\le 2^{r-1}-1$, and $N_2(L(r,c))$ be the number of
$2^n$-periodic binary sequences with  2-error linear complexity
$L(r,c)$.
 Then

{\scriptsize

$N_2(L)=\left\{\begin{array}{l}
\left(\begin{array}{c}2^{n}\\2\end{array}\right)+2^n+1, \ \ \ \  \ \ \ L=0\ \   \\
2^{L-1}(\left(\begin{array}{c}2^{r}\\2\end{array}\right)+2^r+1), \\
 \ \ \ \ \  \ \ \ \ \ \ \ \ \ L=L(r,c), 1\le c\le 2^{r-2}-1, r>2\\
2^{L-1}(\left(\begin{array}{c}2^{r}\\2\end{array}\right)+2^r+1-3\times2^{r+m-3}),  \\
 \ \ \ \ \  \ \ \ \ \ \ \ \ \ L=L(r,c),  c= 2^{r-1}-2^{r-m},  1<m\le r,r\ge2\\
2^{L-1}(\left(\begin{array}{c}2^{r}\\2\end{array}\right)+2^r+1+2^{r-m}-2^{r+m-2}),  \\
 \ \ \ \ \  \ \ \ \ \ \ \ \  L=L(r,c),  c= 2^{r-1}-2^{r-m}+x, \\
\ \ \ \ \  \ \ \ \ \ \ \ \   1<m<r-1, 0<x<2^{r-m-1}, r>3\\
0, \ \ \ \ \ \ \  \ \ \ \ \ \ \ \ \ \  \ \ \ \ \ \ \ \ \ \ \  \ \ \
\mbox{others}
\end{array}\right.$}

\

It is easy to show that Theorem  5.2 is equivalent to the results in
Table 1 and Table 2 of \cite{Kavuluru}.

\

Similarly, based on Theorem  3.1 and Theorem  4.1, the counting functions  for
the number of $2^n$-periodic binary sequences with fixed 3-error
linear complexity can be easily derived as follows.

 \noindent {\bf Theorem  5.3}  Let $L(r,c)=2^n-2^r+c$,  $2\le r\le
n, 1\le c\le 2^{r-1}-1$, and $N_3(L(r,c))$ be the number of
$2^n$-periodic binary sequences with  3-error linear complexity
$L(r,c)$.
 Then

{\scriptsize

$N_3(L)=\left\{\begin{array}{l}
\left(\begin{array}{c}2^{n}\\3\end{array}\right)+\left(\begin{array}{c}2^{n}\\2\end{array}\right)+2^n+1, \ \ \ \  \ \ \ L=0\ \   \\
2^{L-1}(\left(\begin{array}{c}2^{r}\\2\end{array}\right)+1-3\times2^{r+m-3}),  \\
 \ \ \  \ \ \ \ \  L=L(r,c),  c= 2^{r-1}-2^{r-m},  1<m\le r,2\le r\le3\\
2^{L-1}(\left(\begin{array}{c}2^{n}\\3\end{array}\right)+\left(\begin{array}{c}2^{r}\\2\end{array}\right)+2^r+1), \\
 \ \ \ \ \  \ \ \ \ \ \ \ \ \ L=L(r,c), 1\le c\le 2^{r-2}-1, r>2\\
2^{L-1}(\left(\begin{array}{c}2^{r}\\2\end{array}\right)+1-3\times2^{r+m-3}+f(r,m)),  \\
 \ \ \ \ \  \ \ \ \ \ \ \ \ \ L=L(r,c),  c= 2^{r-1}-2^{r-m},  1<m\le r,r>3\\
2^{L-1}(\left(\begin{array}{c}2^{r}\\2\end{array}\right)+1+2^{r-m}-2^{r+m-2}+g(r,m)),  \\
 \ \ \ \ \  \ \ \ \ \ \ \ \  L=L(r,c),  c= 2^{r-1}-2^{r-m}+x, \\
\ \ \ \ \  \ \ \ \ \ \ \ \   1<m<r-1, 0<x<2^{r-m-1}, r>3\\
0, \ \ \ \ \ \ \  \ \ \ \ \ \ \ \ \ \  \ \ \ \ \ \ \ \ \ \ \  \ \ \
\mbox{others}
\end{array}\right.$} where $f(r,m)$ and $g(r,m)$ are defined in Theorem
4.1.

\

According to Table 1 and Table 2 of \cite{Kavuluru}, the numbers of
$2^n$-periodic binary sequences with fixed 3-error linear complexity
for $n=4$ are shown in Table 1.

\begin{center}
\begin{tabular}{ll}
Table 1. $N_3(L(r,c))$  by  \cite{Kavuluru}\\
\hline $L(r,c)$&  $N_3(L(r,c))$\  \\
\hline
 0 & 697   \\
 1 & 697   \\
 2 & 1394   \\
3 & 2788   \\
4 & \underline{5128}   \\
5 & \underline{10704}   \\
6 & \underline{18720}   \\
7 & \underline{30272}   \\
8 & 0   \\
9 & 23808   \\
10 & \underline{22016}   \\
11 & \underline{37888}   \\
12 & 0   \\
13 & 4096   \\
14 & 0   \\
15 & 0  \\
\hline
\end{tabular}
\end{center}

It is well known that the number of all $2^n$-periodic binary
sequences  for $n=4$ is $2^{16}=65536$. However, the summation of
numbers of the right column is much bigger than $65536$. So we
conclude that the counting functions  for the number of
$2^n$-periodic binary sequences with fixed 3-error linear complexity
in \cite{Kavuluru} are not correct.

Specifically, it is easy to check by computer that all underline numbers in  Table 1 are incorrect.

In the case of $L=L(r,c),  c= 2^{r-1}-2^{r-m},  1<m\le r,r\ge2$, which corresponds with the case of $2^n-(2^{n-r_1}+2^{n-r_2})$ in \cite{Kavuluru},  the counting function in \cite{Kavuluru} is wrong.

$L=4,6,7,10$ or 11 in  Table 1 belong to this case.

In the case of $L=L(r,c),  c= 2^{r-1}-2^{r-m}+x,  1<m<r-1, 0<x<2^{r-m-1}, r>3$, which corresponds with the case of $2^n-(2^{n-r_1}+2^{n-r_2})<L<2^n-(2^{n-r_1}+2^{n-r_2-1})$ in \cite{Kavuluru},  the counting function in \cite{Kavuluru} is wrong.

$L=5$  in  Table 1 belongs to this case.

For $n=4, L=5$, we know that $r_1=1, r_2=2$, and $$2^n-(2^{n-r_1}+2^{n-r_2})<L<2^n-(2^{n-r_1}+2^{n-r_2-1})$$

From Theorem 7 in \cite{Kavuluru2008}, we have $N_3(L)=10704$, which is incorrect by computer check.

\

From Theorem  5.3, the numbers of $2^n$-periodic binary sequences
with fixed 3-error linear complexity for $n=4$ are shown in Table 2.
These results have been checked by computer.

\begin{center}
\begin{tabular}{ll}
Table 2. $N_3(L(r,c))$ by Theorem  5.3\\
\hline $L(r,c)$&$N_3(L(r,c))$\\
\hline
 0 & 697   \\
 1 & 697   \\
 2 & 1394   \\
3 & 2788   \\
4 & 2824   \\
5 & 8400   \\
6 & 4384   \\
7 & 2624   \\
8 & 0   \\
9 & 23808   \\
10 & 8704   \\
11 & 5120   \\
12 & 0   \\
13 & 4096   \\
14 & 0   \\
15 & 0  \\
\hline
\end{tabular}
\end{center}

The summation of  numbers of the right column is $2^{16}=65536$.

\section{Conclusion}

 By studying the linear complexity of binary
sequences with period $2^n$, especially the linear complexity will
decline with the superposition of two sequences with same linear
complexity, an approach to construct the complete counting functions
on the $k$-error linear complexity of $2^n$-periodic binary
sequences  was developed. The complete
counting functions on the 2-error linear complexity of
$2^n$-periodic binary sequences were
obtained. The complete counting functions on the 3-error linear
complexity of $2^n$-periodic binary sequences  were also derived.

From Theorem  4.1, we have  counting functions on
the $4$-error  linear complexity for $2^n$-periodic binary sequences with
linear complexity $2^n$. If we have  counting functions on
the $4$-error  linear complexity for $2^n$-periodic binary sequences with
linear complexity less than $2^n$, then we can derive complete counting functions on
the $4$-error  linear complexity, which is our future work.

 \section*{ Acknowledgment}
 The research was supported by
Zhejiang Natural Science Foundation(No.Y1100318, R1090138) and NSAF
(No. 10776077).

\end{document}